%% file: eprint_dpf2013.tex
\documentclass[12pt]{article}
\usepackage{graphicx}
 \usepackage{amsmath}
\makeatletter
\newcommand{\figcaption}{\def\@captype{figure}\caption}
\newcommand{\tabcaption}{\def\@captype{table}\caption}

\newcommand{\Rmnum}[1]{\expandafter\@slowromancap\romannumeral #1@}
\def\hlinewd#1{%
  \noalign{\ifnum0=`}\fi\hrule \@height #1 \futurelet
   \reserved@a\@xhline}
\makeatother

\def\qq{\langle\bar qq\rangle}

\def\GGb{\langle \alpha_sGG\rangle}

\def\GGGb{\langle g_s^3fGGG\rangle}

\def\JJb{\langle g_s^4jj\rangle}
\def\f(s){[(\alpha+\beta)m_c^2-\alpha\beta s]}
\def\non{\\ \nonumber}

\newcommand\pubnumber{DPF2013-2}
\newcommand\pubdate{\today}

\def\napolia{Department of Physics and Engineering Physics\\
University of Saskatchewan, Saskatoon, SK, S7N 5E2, Canada}
\def\napolib{Department of Physics, University of the Fraser Valley, Abbotsford, BC, V2S 7M8, Canada}
\def\napolic{Department of Physics and State Key Laboratory of Nuclear Physics and Technology\\
Peking University, Beijing 100871, China}

\def\Title#1{\begin{center} {\Large #1 } \end{center}}
\def\Author#1{\begin{center}{ \sc #1} \end{center}}
\def\Address#1{\begin{center}{ \it #1} \end{center}}

\newcommand\pubblock{\rightline{\begin{tabular}{l} \pubnumber\\
         \pubdate  \end{tabular}}}
\newenvironment{Abstract}{\begin{quotation}  }{\end{quotation}}
\newenvironment{Presented}{\begin{quotation} \begin{center} 
             PRESENTED AT\end{center}\bigskip 
      \begin{center}\begin{large}}{\end{large}\end{center} \end{quotation}}
\def\Acknowledgments{\bigskip  \bigskip \begin{center} \begin{large}
             \bf ACKNOWLEDGMENTS \end{large}\end{center}}

\input econfmacros.tex

\begin{document}
\begin{titlepage}
\pubblock

\vfill
\Title{QCD Sum Rule Analysis of Heavy Quarkonium Hybrids}
\vfill
\Author{Wei Chen, R. T. Kleiv, and T. G. Steele}
\Address{\napolia}
\Author{B. Bulthuis, D. Harnett, J. Ho, and T. Richards}
\Address{\napolib}
\Author{Shi-Lin Zhu} 
\Address{\napolic}

\vfill
\begin{Abstract}
We have studied charmonium and bottomonium hybrid states with various $J^{PC}$ quantum 
numbers in QCD sum rules. At leading order in $\alpha_s$, the two-point correlation functions have 
been calculated up to dimension six including the tri-gluon condensate and four-quark condensate. 
After performing the QCD sum rule analysis, we have confirmed that the dimension six condensates 
can stabilize the hybrid sum rules and allow  reliable mass predictions. We have updated the mass 
spectra of the charmonium and bottomonium hybrid states and identified that the negative-parity states 
with $J^{PC}=(0, 1, 2)^{-+}, 1^{--}$ form the lightest hybrid supermultiplet while the positive-parity states 
with $J^{PC}=(0, 1)^{+-}, (0, 1, 2)^{++}$ belong to a heavier hybrid supermultiplet. 
\end{Abstract}
\vfill
\begin{Presented}
DPF 2013\\
The Meeting of the American Physical Society\\
Division of Particles and Fields\\
Santa Cruz, California, August 13--17, 2013\\
\end{Presented}
\vfill
\end{titlepage}
\def\thefootnote{\fnsymbol{footnote}}
\setcounter{footnote}{0}

\section{Introduction}
Many new charmonium-like hadron states have been observed at B-factories since 2003~\cite{2011-Brambilla-p1534-1534,
2010-Nielsen-p41-83,2007-Rosner-p12002-12002}. The masses and decay features of 
these states are not consistent with the predictions of the potential model and hence they are considered as candidates of exotic hadron states. 
Some of these states have been interpreted by some authors as charmonium hybrids such as X(3872)~\cite{2003-Close-p210-216}, 
$Y(4260)$~\cite{2005-Zhu-p212-212,2005-Close-p215-222,2007-Klempt-p1-202}, Y(4140)~\cite{2009-Mahajan-p228-228}, and so on. 

Heavy quarkonium hybrids were studied using many methods such as the constituent gluon model~\cite{1978-Horn-p898-898}, 
the flux tube model~\cite{1995-Barnes-p5242-5256}, QCD sum rules~\cite{1985-Govaerts-p215-215,1985-Govaerts-p575-575,
1987-Govaerts-p674-674,1999-Zhu-p31501-31501,2012-Qiao-p15005-15005,2012-Harnett-p125003-125003,2012-Berg-p34002-34002,
2013-Chen-p19-19} 
and lattice QCD~\cite{2012-Liu-p126-126}. In lattice QCD~\cite{2012-Liu-p126-126} and the P-wave quasigluon 
approach~\cite{2008-Guo-p56003-56003}, the heavy quarkonium hybrids with $J^{PC}=(0, 1, 2)^{-+}, 1^{--}$ were predicted to form the 
lightest hybrid supermultiplet while the states with $J^{PC}=0^{+-}, (1^{+-})^3, (2^{+-})^2, 3^{+-}, (0, 1, 2)^{++}$ formed a higher hybrid supermultiplet. 

Heavy quarkonium hybrids were originally studied in Refs.~\cite{1985-Govaerts-p215-215,1985-Govaerts-p575-575,1987-Govaerts-p674-674} by Govaerts \textit{et al.}. 
They performed the QCD sum-rule analyses considering the perturbative and 
the dimension four gluon condensate contributions, which led to unstable hybrid sum rules and unreliable mass predictions for the
$J^{PC}=0^{-+}, 0^{+-}, 1^{-+}, 1^{--}, 2^{-+}$ channels. Recently, the $J^{PC}=1^{--}$~\cite{2012-Qiao-p15005-15005}, $1^{++}$~\cite{2012-Harnett-p125003-125003} and $0^{-+}$~\cite{2012-Berg-p34002-34002} channels have been re-analyzed by including the dimension six condensate contributions to the two-point correlation functions. The results showed that the tri-gluon condensate contributions can stabilize the hybrid sum rules and allow reliable mass predictions. 

In this contribution, we review our work in Ref.~\cite{2013-Chen-p19-19} where we extend the calculation of the correlation functions of heavy quarkonium hybrid 
operators with various $J^{PC}$ quantum numbers to include QCD condensates up to dimension six. After evaluating the spectral densities, we re-analyze both the 
charmonium and bottomonium hybrid channels and update their mass spectra. 

\section{Two-point Correlation Functions}\label{sec:QSR}
In the framework of QCD sum rules~\cite{1979-Shifman-p385-447,1985-Reinders-p1-1}, 
we consider the following interpolating currents with various $J^{PC}$ quantum numbers
\begin{eqnarray}
\nonumber
J_{\mu}&=&g_s\bar Q\frac{\lambda^a}{2}\gamma^{\nu}G^a_{\mu\nu}Q,~~~~~~~J^{PC}=1^{-+}, 0^{++},
\\
J_{\mu}&=&g_s\bar Q\frac{\lambda^a}{2}\gamma^{\nu}\gamma_5G^a_{\mu\nu}Q,~~~~J^{PC}=1^{+-}, 0^{--}, \label{currents}
\non
J_{\mu\nu}&=&g_s\bar Q\frac{\lambda^a}{2}\sigma_{\mu}^{\alpha}\gamma_5 G^a_{\alpha\nu}Q,~~~~J^{PC}=2^{-+}, 1^{++}, 1^{-+}, 0^{-+}\,,
\end{eqnarray}
in which $Q$ represents a heavy quark ($c$ or $b$), $g_s$ is the strong coupling, $\lambda^a$ are the Gell-Mann matrices and $G^a_{\mu\nu}$ is the gluon field strength. By replacing $G^a_{\mu\nu}$ with $\tilde G^a_{\mu\nu}=\frac{1}{2}\epsilon_{\mu\nu\alpha\beta}G^{\alpha\beta,a}$,
we can also obtain the corresponding operators with opposite parity. Using these hybrid operators, we study the two-point correlation functions 
\begin{eqnarray}
\Pi_{\mu\nu}(q)= i\int
d^4x \,e^{iq \cdot x}\,\langle0|T[J_{\mu}(x)J_{\nu}^{\dag}(0)]|0\rangle, \label{equ:Pi}
\end{eqnarray}
where $J_{\mu}(x)$ is the interpolating current in Eq.~\eqref{currents}. 

At the hadron level, the correlation functions can be described using the dispersion relation 
\begin{eqnarray}
\Pi(q^2)=(q^2)^N\int_{4m^2}^{\infty}\frac{\rho(s)}{s^N(s-q^2-i\epsilon)}ds+\sum_{n=0}^{N-1}b_n(q^2)^n,
\end{eqnarray}
where $\rho(s)$ is the spectral density
\begin{eqnarray}
\rho(s)\equiv\sum_n\delta(s-m_n^2)\langle0|J_{\mu}|n\rangle\langle n|J_{\mu}^{\dagger}|0\rangle
=f_X^2m_X^8\delta(s-m_X^2)+ \mbox{continuum},  \label{Phenrho}
\end{eqnarray}
and $f_X$ is the coupling constant and $m_X$ is the mass of the ground state. The correlation function can also be calculated at the quark-gluon level via 
the OPE (operator product expansion) method. Up to dimension six, the correlation functions and spectral densities can be 
expressed as the sum of a perturbative term and the various QCD condensates such as the gluon condensate, tri-gluon condensate and the 
four-quark condensate. One can consult Ref.~\cite{2013-Chen-p19-19} for the detailed expressions for the spectral densities.

After performing the Borel transform, we establish the sum rules for the hadron mass by equating the correlation functions obtained at 
both the hadron level and the quark-gluon level 
\begin{eqnarray}
\mathcal{L}_{k}\left(s_0, M_B^2\right)=f_X^2 m_X^{8+2k}e^{-m_X^2/M_B^2}=\int_{4m^2}^{s_0}ds\,s^k\,\rho(s)\,e^{-s/M_B^2}\,,
\label{sumrule}
\end{eqnarray}
where $s_0$ and $M_B$ are the continuum threshold and the Borel mass, respectively. The mass of the lowest-lying hybrid state can be extracted as 
\begin{eqnarray}
m_X^2=\frac{\mathcal{L}_{1}\left(s_0\,, M_B^2\right)}{\mathcal{L}_{0}\left(s_0\,, M_B^2\right)}\,.
\label{mass}
\end{eqnarray}

\section{QCD sum rule analysis}\label{sec:NA}
To perform the numerical analysis, we use the following values of the heavy quark masses and the various 
condensates~\cite{2009-Chetyrkin-p74010-74010, 2012-Narison-p259-263, 2010-Narison-p559-559,2007-Kuhn-p192-215}: 
$m_c(\mu=m_c)=\overline m_c=(1.28\pm 0.02)~\mbox{GeV}, m_b(\mu=m_b)=\overline m_b=(4.17\pm 0.02)~\mbox{GeV}, 
\GGb=(7.5\pm2.0)\times 10^{-2}~\mbox{GeV}^4$, $\GGGb=-(8.2\pm1.0)~\mbox{GeV}^2\GGb, \qq=-(0.23\pm0.03)^3~\mbox{GeV}^3, 
\JJb=-\frac{4}{3}g_s^4\qq^2$. 
The stability of the mass sum rules requires suitable working regions of the continuum threshold $s_0$ and Borel mass $M_B^2$. 
We study the convergence of the OPE series and the pole contribution to obtain the Borel window. The requirement of the OPE 
convergence results in a lower bound on $M_B^2$ while the constraints of the pole contribution leads to an upper bound. 

For the charmonium hybrid $\bar cGc$ systems, the dominant power correction comes from the gluon condensate $\GGb$. However, the 
dimension six condensates $\GGGb$ and $\JJb$ are also very important to the hybrid sum rules. They can improve the convergence of the 
OPE series and stabilize the mass sum rules~\cite{2012-Qiao-p15005-15005,2012-Harnett-p125003-125003,2012-Berg-p34002-34002,2013-Chen-p19-19}. 
To obtain a lower bound on the Borel mass, we require that the gluon condensate be less than one third of the perturbative term contribution 
and that the tri-gluon condensate be less than one third of the gluon condensate contribution. An upper bound on $M_B^2$ is determined by 
requiring the pole contribution be larger than $10\%$. For the exotic channel $J^{PC}=1^{-+}$, the Borel window is then determinded as 
$4.6$ GeV$^2$ $\leq M_B^2\leq 6.5$ GeV$^2$. 

\begin{figure}[htb]
\centering
\begin{tabular}{lr}
\scalebox{0.55}{\includegraphics{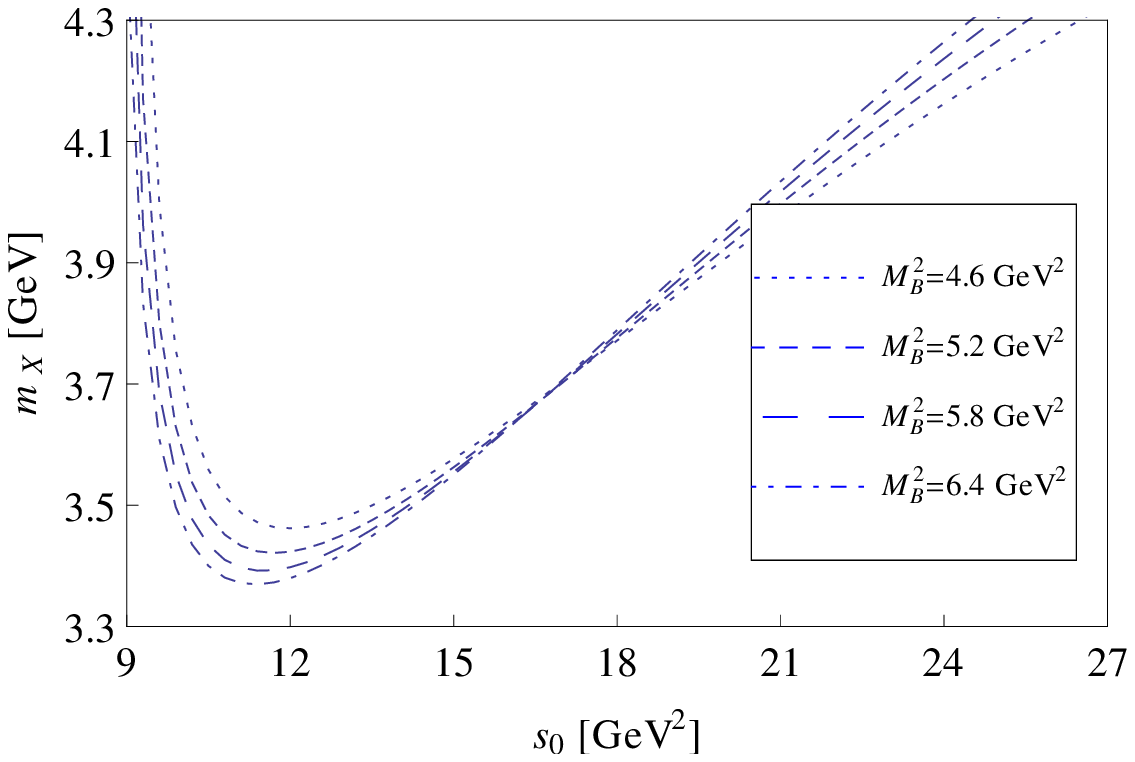}}&
\scalebox{0.55}{\includegraphics{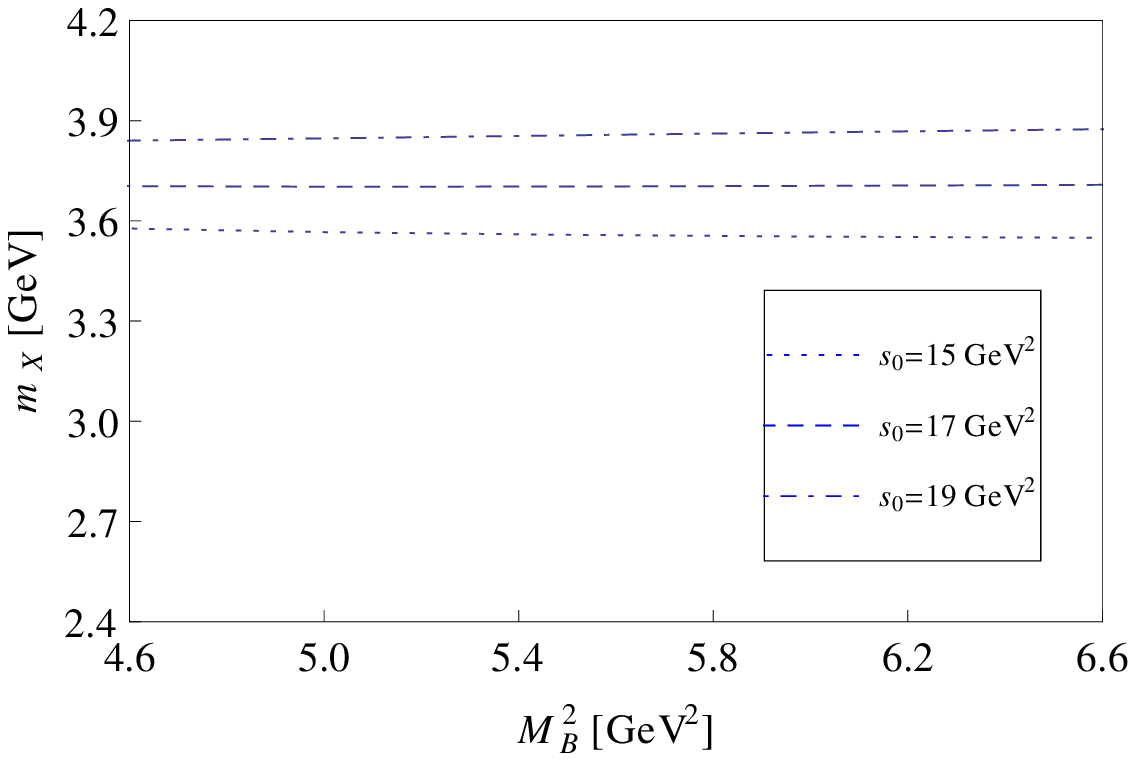}}
\end{tabular}
\caption{The variations of $m_X$ with $s_0$ and $M_B^2$ for the $1^{-+}$ charmonium hybrid.}\label{fig1-+cc}
\end{figure}

In this Borel window, we study the variation of the extracted hybrid mass $m_X$ with $s_0$ for the $J^{PC}=1^{-+}$ channel in the LHS of 
Fig.~\ref{fig1-+cc}. This figure gives us the most suitable value of the continuum threshold $s_0=17$ GeV$^2$, around which the variation 
of $m_X$ with $M_B^2$ is minimum. In the RHS of Fig.~\ref{fig1-+cc}, we study the Borel curve which describes the variation of $m_X$ with 
$M_B^2$. The Borel curve is very stable in the Borel window. Finally, we extract the mass of the $1^{-+}$ charmonium hybrid as $m_X=3.70$ 
GeV, which is about $0.5$ GeV lower than the lattice result in Ref.~\cite{2012-Liu-p126-126}.

Following the same procedure, we extract masses of the other charmonium hybrids as summarized in Table~\ref{table1} along with the 
corresponding Borel windows, threshold values and pole contributions. Only errors from the uncertainties in the charm quark mass 
and the condensates are taken into account. We do not consider other possible error sources such as truncation of the OPE series, 
the uncertainty of the threshold value $s_0$ and the variation of Borel mass $M_B$.

\begin{table}[t]
\begin{center}
\begin{tabular}{cccccc}
\hlinewd{.8pt}
& $J^{PC}$ & $s_0(\mbox{GeV}^2)$&$[M^2_{\mbox{min}}$,$M^2_{\mbox{max}}](\mbox{GeV}^2)$&$m_X$\mbox{(GeV)}&PC(\%)\\
\hline
& $1^{--}$   & 15  & $2.5\sim4.8 $& $3.36\pm0.15$& 18.3\\
& $0^{-+}$   & 16  & $5.6\sim7.0 $& $3.61\pm0.21$& 15.4\\
& $1^{-+}$   & 17  & $4.6\sim6.5 $& $3.70\pm0.21$& 18.8\\
& $2^{-+}$   & 18  & $3.9\sim7.2 $& $4.04\pm0.23$& 26.0
\vspace{5pt}\\
& $0^{+-}$   & 20  & $6.0\sim7.4 $& $4.09\pm0.23$& 15.5\\
& $2^{++}$   & 23  & $3.9\sim7.5$ & $4.45\pm0.27$& 21.5\\
& $1^{+-}$   & 24  & $2.5\sim8.4 $& $4.53\pm0.23$& 33.2\\
& $1^{++}$   & 30  & $4.6\sim11.4$& $5.06\pm0.44$& 30.4\\
& $0^{++}$   & 34  & $5.6\sim14.6$& $5.34\pm0.45$& 36.3
\vspace{5pt}\\
& $0^{--}$    & 35  & $6.0\sim12.3$& $5.51\pm0.50$& 31.0\\
\hline
\hlinewd{.8pt}
\end{tabular}
\caption{Numerical results for the charmonium hybrid states. \label{table1}}
\end{center}
\end{table}

The unstable channels with $J^{PC}=0^{-+}, 0^{+-}, 1^{-+}, 1^{--}, 2^{-+}$ in Refs.~\cite{1985-Govaerts-p215-215,1985-Govaerts-p575-575,
1987-Govaerts-p674-674} are stable in Table~\ref{table1}. Obviously, the dimension six condensate $\GGGb$ and $\JJb$ in the correlation 
functions can stabilize the sum rules and make it possible to extract hybrid masses.
In Table~\ref{table1}, the charmonium hybrids with $J^{PC}=(0, 1, 2)^{-+}, 1^{--}$ lie in the range $3.4\sim3.9$ GeV, which are much lower 
than the other channels. They form the lightest hybrid supermultiplet~\cite{2012-Liu-p126-126,2008-Guo-p56003-56003}. A heavier hybrid supermultiplet in
Refs.~\cite{2012-Liu-p126-126, 2008-Guo-p56003-56003} contains states with $J^{PC}=0^{+-}, (1^{+-})^3, (2^{+-})^2, 3^{+-}, (0, 1, 2)^{++}$.
In Table~\ref{table1}, we obtain five members of this excited hybrid supermultiplet with $J^{PC}=(0, 1)^{+-}, (0, 1, 2)^{++}$.

As in the previous result in Ref.~\cite{2012-Harnett-p125003-125003}, the mass of the $1^{++}$ charmonium hybrid in Table~\ref{table1}
is around $5.06$ GeV, which is much higher than the mass of $X(3872)$. It seems that a pure hybrid interpretation of this meson is precluded. 

We perform the same analysis as described above by replacing $m_c$ with $m_b$ in the spectral densities to study the bottomonium hybrid systems. The numerical results are then collected in Table~\ref{table2}. It shows that the masses of the four bottomonium hybrid states with $J^{PC}=(0, 1, 2)^{-+}, 1^{--}$ are about $9.7\sim 9.9$ GeV, which form the lightest bottomonium supermultiplet.
\begin{table}[t]
\begin{center}
\begin{tabular}{cccccc}
\hlinewd{.8pt}
& $J^{PC}$ & $s_0(\mbox{GeV}^2)$&$[M^2_{\mbox{min}}$,$M^2_{\mbox{max}}](\mbox{GeV}^2)$&$m_X$\mbox{(GeV)}&PC(\%)\\
\hline
& $1^{--}$    & 105  & $11\sim17 $& $9.70\pm0.12$&  17.2\\
& $0^{-+}$   & 104  & $14\sim16 $& $9.68\pm0.29$& 17.3\\
& $1^{-+}$   & 107  & $13\sim19 $& $9.79\pm0.22$& 20.4\\
& $2^{-+}$   & 105  & $12\sim19$& $9.93\pm0.21$& 21.7
\vspace{5pt}\\
& $0^{+-}$   & 114  & $14\sim19 $& $10.17\pm0.22$& 17.6\\
& $2^{++}$   & 120  & $12\sim20$ & $10.64\pm0.33$& 19.7\\
& $1^{+-}$   & 123  & $10\sim21 $& $10.70\pm0.53$& 28.5\\
& $1^{++}$  & 134  & $13\sim27$& $11.09\pm0.60$& 27.7\\
& $0^{++}$  & 137  & $13\sim31$& $11.20\pm0.48$& 30.0
\vspace{5pt}\\
& $0^{--}$    & 142  & $14\sim25$& $11.48\pm0.75$& 24.1\\
\hline
\hlinewd{.8pt}
\end{tabular}
\caption{Numerical results for the bottomonium hybrid states.\label{table2}}
\end{center}
\end{table}

\section{Summary}\label{sec:Summary}
We have reviewed our work in Ref.~\cite{2013-Chen-p19-19} where we studied the charmonium and bottomonium hybrids 
with various quantum numbers in QCD sum rules. At leading order in $\alpha_s$, the two-point correlation functions 
and the spectral densities are calculated including the dimension six tri-gluon condensate and the four-quark condensate. 

For both the charmonium and bottomonium hybrid systems, the numerical analyses show that the gluon condensates 
$\GGb$ are the dominant power corrections to the correlation functions. However, the dimension six condensates 
$\GGGb$ and $\JJb$ can improve the OPE convergence, stabilize the mass sum rules and thus make the mass predictions 
reliable. For the $J^{PC}=1^{++}$ charmonium hybrid channel, the mass is extracted around $5.06$ GeV, which is 
much higher than the mass of $X(3872)$. This precludes a pure hybrid interpretation of this meson. 

In our results, the negative-parity hybrids with $J^{PC}=(0, 1, 2)^{-+}, 1^{--}$ are much lighter than the other states. 
In other words, these states form the lightest hybrid supermultiplet.  At the same time, the positive-parity states with 
$J^{PC}=(0, 1)^{+-}, (0, 1, 2)^{++}$ belong to a heavier hybrid supermultiplet. The hybrid with $J^{PC}=0^{--}$ is 
the heaviest, which implies a very different gluonic excitation.

\Acknowledgments
This project was supported by the Natural Sciences and Engineering
Research Council of Canada (NSERC). S.L.Z. was supported by the
National Natural Science Foundation of China under Grants
11075004, 11021092, 11261130311 and Ministry of Science and
Technology of China (2009CB825200).


\end{document}

%% file: econfmacros.tex
\def\beq{\begin{equation}}
\def\eeq#1{\label{#1}\end{equation}}
\def\eeqn{\end{equation}}

\def\beqa{\begin{eqnarray}}
\def\eeqa#1{\label{#1}\end{eqnarray}}
\def\eeqan{\end{eqnarray}}

\let\bar=\overbar

\def\Dslash{\not{\hbox{\kern-4pt $D$}}}
\def\dslash{\not{\hbox{\kern-2pt $\del$}}}

\def\msb{{\bar{\ssstyle M \kern -1pt S}}}

